\documentclass[%
reprint,  
amsmath,amssymb,
aps,
]{revtex4-2}

\usepackage[pdftex]{graphicx}
\usepackage{bm}

\usepackage{mathtools}
\usepackage{physics2}
\usephysicsmodule{ab, ab.legacy, op.legacy}
\usepackage{fixdif, derivative}
\usepackage{booktabs}
\usepackage{comment}

\newcommand{\jointarticle}{\cite{ishii2025Qualitatively}}
\newcommand{\rmsub}[2]{#1_\mathrm{#2}}
\newcommand{\rmsup}[2]{#1^\mathrm{#2}}
\newcommand{\din}[1]{d^\mathrm{in}\if\relax\detokenize{#1}\relax\else_{#1}\fi}
\newcommand{\dout}[1]{d^\mathrm{out}\if\relax\detokenize{#1}\relax\else_{#1}\fi}

\begin{document}


\title{%
Degree heterogeneity shapes escape mechanisms in networks of diffusively coupled bistable elements
}

\author{Hidemasa Ishii}
\email{hidemasaishii1997@g.ecc.u-tokyo.ac.jp}
\author{Hiroshi Kori}%
\affiliation{%
Department of Complexity Science and Engineering, Graduate School of Frontier Sciences, the University of Tokyo, Chiba 277-8561, Japan
}%


\date{\today}

\begin{abstract}
For fully connected populations of diffusively coupled bistable elements, we identified three qualitatively distinct mechanisms of noise-induced escape as coupling strength varies [H. Ishii and H. Kori, arXiv:2512.01388 (2025)].
Here we generalize these results to a class of networked systems and demonstrate that degree heterogeneity (i.e., variability in node degree) shapes escape mechanisms alongside coupling strength.
In applied contexts, networks of noisy bistable elements provide a minimal conceptual framework for understanding abrupt state transitions in complex systems.
Theoretically, a quantitative approach to escape is challenging because nonlinearity, network interactions, and dynamical noise jointly shape the collective dynamics.
We extend the analytical framework developed for the fully connected model to a class of networked systems based on the annealed network approximation.
We derive three effective one-dimensional descriptions of collective escape dynamics.
We validate our theoretical predictions for mean escape times by direct numerical simulations.
Our analysis reveals that the validity and quantitative behavior of the reduced descriptions depend on degree heterogeneity in addition to coupling strength.
This work extends the classification of escape mechanisms to networked bistable elements.
Furthermore, our analytical framework provides tools for understanding synergistic phenomena arising from the interplay of nonlinearity, diffusive coupling, and dynamical noise.
\end{abstract}

\maketitle


\section{Introduction}
Diverse complex systems often exhibit switching behavior between typical states, also known as regime shifts and tipping events.
Bistable models have served as a minimal conceptual framework to investigate such abrupt changes from epilepsy~\cite{kalitzin2010Stimulationbased} to social uprisings~\cite{brummitt2015Coupled}.
In such a framework, each stable state corresponds to a single typical state, and external forcings induce state transitions~\cite{ashwin2012Tipping}.
In particular, when external forcings are stochastic, the switching behavior is called noise-induced escape~\cite{ashwin2017Fast}.
Since noise-induced escape necessitates both the system's bistability and noise, it represents a phenomenon that is both practically relevant and theoretically interesting.

Many complex systems consist of multiple dynamical elements interacting on a network.
Each element, not only the system as a whole, often exhibits bistability and switching behavior.
For instance, an epileptic brain has been conceptually modeled as a network of bistable brain regions, each of which transitions between normal and seizure states~\cite{benjamin2012Phenomenological,creaser2020Dominolike}.
The Earth climate system consists of interacting tipping (i.e., bistable) elements, such as rainforests and ice sheets~\cite{wunderling2021Interacting}.
Accordingly, this research studies collective escape dynamics of networked populations, wherein each noisy bistable element transitions from one state to the other.
To understand the dynamics, one must analyze the interplay of the system's nonlinearity, networked interaction, and dynamical noise.

Pioneered by Kramers in 1940~\cite{kramers1940Brownian}, the literature on noise-induced escape has extended its scope to populations of bistable elements, where the escape of one element affects the dynamics of others through coupling.
Previous studies~\cite{frankowicz1982Stochastic,berglund2007Metastability,berglund2007Metastabilitya,ashwin2017Fast,creaser2018Sequential,ashwin2018Sequential,ashwin2023Quasipotentials,ishii2024Diffusivea} have analyzed the influence of changes in the coupling strength on escape dynamics, which are often characterized by the mean escape time --- i.e., the expected time until the system escapes.
Coupling is often assumed to be diffusive~\cite{frankowicz1982Stochastic,berglund2007Metastability,berglund2007Metastabilitya,ashwin2017Fast,creaser2018Sequential,ashwin2023Quasipotentials}, facilitating synchronization among elements.
This class of systems, i.e., populations of diffusively coupled bistable elements, models networks of bistable electrochemical cells~\cite{kouvaris2016SelfOrganized,kouvaris2017Stationary}.
Such systems can also be seen as a variant of consensus models~\cite{olfati-saber2007Consensus} where each agent exhibits bistability.

Despite the accumulated literature on noise-induced escape, we identify a gap in our knowledge of collective escape dynamics.
Systems of diffusively coupled bistable elements exhibit bifurcations with respect to the system-wide coupling strength.
Accordingly, many existing studies~\cite{frankowicz1982Stochastic,berglund2007Metastability,berglund2007Metastabilitya,ashwin2017Fast,creaser2018Sequential,ashwin2018Sequential,ashwin2023Quasipotentials} have focused on bifurcations of the noise-free systems.
In particular, the following parameter regimes have been distinguished:
the weak-coupling regime, where all the fixed points of the uncoupled system exist;
the intermediate ``slow-domino'' regime, where fixed points disappear through saddle-node bifurcations;
and the synchronized or ``fast-domino'' regime, which is beyond the largest bifurcation point~\cite{berglund2007Metastability,ashwin2017Fast}.
There are three limitations in this bifurcation-based approach.
Firstly, it cannot explain the effect of varying coupling strength in the last regime.
In our previous study, we numerically showed that the mean escape time first declines, then increases, and then reaches an asymptotic value as the coupling strength is increased~\cite{ishii2024Diffusivea}.
Since this non-monotonic behavior is observed in the fast-domino regime where no bifurcation occurs, one must consider other factors to explain it.
Secondly, it is often infeasible to analyze bifurcations of large or networked systems.
Indeed, most previous works have either considered small systems with a few nodes~\cite{ashwin2017Fast,creaser2018Sequential,creaser2020Sequential,ashwin2023Quasipotentials} or conducted numerical and data-driven analyses of large and networked systems~\cite{benjamin2012Phenomenological,creaser2020Dominolike,lopes2023Role,ishii2024Diffusivea}.
Last but not least, such an approach offers little insight into the influence of differing network properties on escape dynamics, as far as the system-wide coupling strength is treated as a bifurcation parameter.
To overcome these limitations, it is desirable to develop coarse-grained theoretical approaches that do not rely on detailed bifurcation structure or system-specific information such as the full adjacency matrix.

In this article, we develop a model-reduction approach to noise-induced escape of networked populations of bistable elements with diffusive coupling.
Our central focus is the influence of varying degree heterogeneity $\kappa$, which quantifies the variation in node degree within a network, along with coupling strength $K$.
After introducing the model and defining the mean escape time, we derive effective one-dimensional dynamics that describe collective escape processes using the annealed network approximation.
We perform model reduction in three regimes of coupling strength, which are distinguished according to the validity of the effective dynamics.
Our reduction scheme, which extends the analysis of the accompanying Letter~\jointarticle{}, is applicable to other diffusively coupled nonlinear random dynamical systems.
We validate our theory by comparing predicted mean escape times with direct numerical simulations.
The analysis reveals that degree heterogeneity plays a crucial role in shaping escape dynamics alongside coupling strength.
In our model, increasing degree heterogeneity attenuates the dependence of escape dynamics on the coupling strength by enhancing the influence of high-degree nodes on the collective dynamics.

\section{Model and mean escape time}
We analyze a system consisting of $N$ identical bistable nodes that interact on a network represented by the adjacency matrix $A$.
The model is governed by the following stochastic differential equations (SDEs):
\begin{subequations} \label{eq:original.model}
  \begin{align}
  \dot{x}_i =& f(x_i) + \frac{K}{\din{i}} \sum_{j=1}^N A_{ij} \pab{x_j - x_i} + \sqrt{2D} \xi_i,  \\
  f(x) \coloneq& -x \pab{x - r} \pab{x - 1},
\end{align} \end{subequations}
where $K$ is the coupling strength, $\din{i}$ denotes the indegree of node $i$, $D$ is the noise strength, and $\xi_i$ represents mutually independent white Gaussian noise satisfying $\aab{\xi_i(t)} = 0$ and $\aab{\xi_i(t) \xi_j(t+\tau)} = \delta_{ij} \delta(\tau)$.
The $(i, j)$ entry of $A$ (i.e., $A_{ij}$) is $1$ if an edge from $j$ to $i$ exists and $0$ otherwise.
The local flow $f(x)$ induces bistability of each node in the absence of coupling and noise.
In the remainder, $U(x)$ denotes the associated double-well potential that satisfies $f(x) = -\d U / \d x$.
The system $\dot{x} = f(x)$ has stable fixed points at $x = 0$ and $1$ and an unstable fixed point at $x = r$.
Throughout this work, we assume $0 < r \ll 0.5$, in which case $x = 1$ is much more stable than $x = 0$.
In the following, we refer to $x = 0$ and $x = 1$ as the ``background'' and ``active'' states, respectively.
This strong asymmetry allows us to focus on escape from the background to the active state while neglecting the opposite transition.
While a small $r$ is crucial for numerically measuring mean escape times, our model reduction does not rely on this assumption.

In our model, the coupling term is normalized by the indegree $\din{i}$.
This ensures that the coupling strength $K$ quantifies the influence of coupling relative to the local bistable dynamics regardless of node degree.
Without normalization, the dynamics of high-degree nodes would be dominated by coupling, whereas coupling would have little influence on low-degree nodes.
In social contexts, this normalization corresponds to assuming that the influence of each friend is inversely proportional to the number of friends.
In chemical experiments, such normalization would be realized when the volume of each reactor scales with its indegree.

We define the mean escape time as follows.
First, the system is initialized to the collective background state, $\bm{x}_0 = (0, \dots, 0)^\top$.
Then, the first escape time of node $i$ is defined as
\begin{gather}
  \tau_i \coloneq \inf\Bab{t > 0 \,\Big|\, x_i(t) \geq \xi \text{ given } x_i(0) = 0},
\end{gather}
where $\xi$ is a fixed threshold between the background and active states.
We use $\xi = 0.5$ in the following, but this choice does not affect our results as long as $\xi$ is not close to $r$ or $1$.
Since $\tau_i$ is defined for each node, we first take an average over nodes within a given network.
We then take the expectation over noise realizations of this node-averaged quantity, defining it as the mean escape time $\bar{\tau}$:
\begin{gather}
  \bar{\tau} \coloneq \aab{\frac{1}{N} \sum_{i=1}^N \tau_i}.
\end{gather}

For a one-dimensional gradient system, the following formula for the mean first passage time~\cite[Section~5.5]{gardiner2009Stochastic} is known.
Specifically, consider a stochastic process governed by $\dot{x} = -V'(x) + \sqrt{2D} \xi$ on $(-\infty, \xi)$, with a reflecting boundary at $x = -\infty$ and an absorbing boundary at $x = \xi$.
Here, $V(x)$ denotes the associated potential.
The mean first passage time $T$ for a particle starting at $x = 0$ to exit through $x = \xi$ is given by
\begin{gather} \label{eq:mfpt}
  T = \frac{1}{D} \int_0^\xi \d y \int_{-\infty}^y \d z \exp\pab{-\frac{V(y) - V(z)}{D}}.
\end{gather}
Replacing $V(x)$ by the double-well potential $U(x)$ allows us to calculate the mean escape time of a single element, which coincides with the mean escape time of the uncoupled population (i.e., $K = 0$):
\begin{gather} \label{eq:T0}
  T_0 = \frac{1}{D} \int_0^\xi \d y \int_{-\infty}^y \d z \exp\pab{-\frac{U(y) - U(z)}{D}}.
\end{gather}
This formula of $T_0$ simplifies to the Eyring-Kramers law~\cite{berglund2013Kramers} in the low-noise limit~\cite{ishii2024Diffusivea}.

\section{Annealed network approximation}
To facilitate further analysis, we employ the annealed network approximation~\cite{boguna2009Langevin,pastor-satorras2015Epidemic} to simplify the model.
This approximation replaces the actual network $A$ by its statistical average $\aab{A}$, where each entry $\aab{A_{ij}}$ represents the expected number of edges from node $j$ to node $i$.
More precisely, we replace $A_{ij}$ by
\begin{gather} \label{eq:ana}
  \aab{A_{ij}} = \dout{j} \frac{\din{i}}{N \aab{\din{}}}.
\end{gather}
In other words, this approximation replaces the network $A$ by the corresponding weighted, fully connected graph $\aab{A}$.
Details are provided in Appendix~\ref{app:ana}.
In this way, the degree of each node is preserved in the sense that it equals the total weight of the node.
That is, it holds that $\sum_{i=1}^N \aab{A_{ij}} = \dout{j}$ and $\sum_{j=1}^N \aab{A_{ij}} = \din{i}$.
This approximation is justified when the underlying network rewires on a time scale much faster than the dynamics on it, while conserving each node's degree~\cite{boguna2009Langevin}.
Put differently, it is a rather strong approximation when analyzing dynamical processes on a static network~\cite{pastor-satorras2015Epidemic}.
Accordingly, our theoretical predictions based on this approximation will be validated numerically.
We distinguish this procedure from the degree-based mean-field approach~\cite{pastor-satorras2015Epidemic,kundu2022Meanfield}, which is characterized by introducing state variables $\Psi_d$ that represent the state of nodes with degree $d$.
While the degree-based mean-field approach typically reduces the system's degrees of freedom, the number of state variables remains $N$ under the annealed network approximation.

Replacing $A_{ij}$ by $\aab{A_{ij}}$ [Eq.~\eqref{eq:ana}] in our model [Eq.~\eqref{eq:original.model}], we obtain the following approximate model:
\begin{gather} \label{eq:approx.model}
  \dot{x}_i = f(x_i) + K \sum_{j=1}^N \rho_j \pab{x_j - x_i} + \sqrt{2D} \xi_i,
\end{gather}
where we defined the normalized outdegree
\begin{gather}
  \rho_j \coloneq \frac{\dout{j}}{N \aab{\dout{}}},
\end{gather}
which is normalized as $\sum_{j=1}^N \rho_j = 1$.
Introducing the (out)degree-weighted mean field
\begin{gather}
  \Theta(t) \coloneq \sum_{i=1}^N \rho_i x_i(t),
\end{gather}
the approximate model can be rewritten as
\begin{gather}
  \dot{x}_i = f(x_i) + K \pab{\Theta - x_i} + \sqrt{2D} \xi_i,
\end{gather}
which is similar to the globally coupled model analyzed in the companion Letter~\jointarticle{}, motivating us to extend its analysis to networks.
The theoretical analysis below focuses on the approximate model.
We test its validity through numerical simulations of the original model.

We define the deviation of node $i$ from $\Theta$:
\begin{gather}
  y_i \coloneq x_i - \Theta,
\end{gather}
which satisfies $\sum_{i=1}^N \rho_i y_i = 0$.
Assuming small $y_i$, we expand $f(x_i) = f(\Theta + y_i)$ around $\Theta$ to the second order, obtaining the following evolution equation of $\Theta$:
\begin{gather} \label{eq:theta.dyn}
  \dot{\Theta} = \sum_{i=1}^N \rho_i \dot{x}_i 
    \approx f(\Theta) + \frac{f''(\Theta)}{2} Z + \sqrt{2D} \hat{\eta}_\Theta,
\end{gather}
where we defined the weighted mean square deviation
\begin{gather} \label{eq:Z}
  Z(t) \coloneq \sum_{j=1}^N \rho_j \bab{y_j(t)}^2.
\end{gather}
When the network is degree-homogeneous, i.e., $\rho_j \equiv 1 / N$, the quantity $Z$ corresponds to the variance of $y_j$.
We also introduced the effective white Gaussian noise
\begin{gather}
  \hat{\eta}_\Theta(t) \coloneq \sum_{i=1}^N \rho_i \xi_i(t),
\end{gather}
which has the following statistical properties:
\begin{subequations}
  \begin{align}
    \aab{\hat{\eta}_\Theta(t)} =& 0,  \\
    \aab{\hat{\eta}_\Theta(t) \hat{\eta}_\Theta(t+\tau)} =& \delta(\tau) \frac{\kappa}{N}.
  \end{align}
\end{subequations}
Here, we used the network's (out)degree heterogeneity $\kappa$, defined by
\begin{gather} \label{eq:kappa}
  \kappa \coloneq \frac{\aab{\pab{\dout{}}^2}}{\aab{\dout{}}^2} = N \sum_{i=1}^N \rho_i^2.
\end{gather}
Literally, $\kappa$ quantifies the heterogeneity of outdegree within the network.
It holds that $1 \leq \kappa < N$, and the minimum is achieved when all nodes have the same outdegree as in fully connected graphs.
Linearizing $f(x_i)$ around $\Theta$, we obtain the following dynamics of $y_i$:
\begin{gather} \label{eq:yi.dyn}
  \dot{y}_i = -\bab{K - f'(\Theta)} y_i + \sqrt{2D} \hat{\eta}_i,
\end{gather}
where we introduced effective white Gaussian noise
\begin{gather}
  \hat{\eta}_i(t) \coloneq \xi_i(t) - \hat{\eta}_\Theta(t),
\end{gather}
which has the following statistical properties:
\begin{subequations}
\begin{align}
  \aab{\hat{\eta}_i(t)} =& 0,  \\
  \label{eq:corr.etahati}
  \aab{\hat{\eta}_i(t) \hat{\eta}_j(t+\tau)} =& \delta(\tau) \pab{\delta_{ij} - \rho_i - \rho_j + \frac{\kappa}{N}},  \\
  \aab{\hat{\eta}_i(t) \hat{\eta}_\Theta(t+\tau)} =& \delta(\tau) \pab{\rho_i - \frac{\kappa}{N}}.
\end{align}
\end{subequations}
To summarize, the annealed network approximation yields the approximate model [Eq.~\eqref{eq:approx.model}], which in turn gives the evolution equations of $\Theta$ [Eq.~\eqref{eq:theta.dyn}] and $y_i$ [Eq.~\eqref{eq:yi.dyn}] for sufficiently small $y_i$.
We highlight that, for directed networks, only the heterogeneity in outdegree affects escape dynamics under the annealed network approximation.
The normalization by indegree in our model [Eq.~\eqref{eq:original.model}] cancels the contribution of indegree heterogeneity.

The derived dynamics indicate how degree heterogeneity $\kappa$ contributes to the collective escape.
In the strong-coupling limit $K \to \infty$, the deviation dynamics [Eq.~\eqref{eq:yi.dyn}] suggest $y_i \to 0$, implying that the following one-dimensional dynamics of $\Theta$ fully represents the evolution of the network:
\begin{gather}
  \dot{\Theta} = f(\Theta) + \sqrt{2D} \sqrt{\frac{\kappa}{N}} \eta_\Theta,
\end{gather}
where $\eta_\Theta(t)$ is an effective white Gaussian noise with the same statistical properties as $\xi_i$, i.e., $\aab{\eta_\Theta(t)} = 0$ and $\aab{\eta_\Theta(t) \eta_\Theta(t + \tau)} = \delta(\tau)$.
Using the mean first passage time formula [Eq.~\eqref{eq:mfpt}], the mean escape time is given by
\begin{gather} \label{eq:T.infty}
  T_\infty\pab{\frac{\kappa}{N}} = \frac{N}{\kappa D} \int_0^\xi \d y \int_{-\infty}^y \d z \exp\pab{-\frac{N \bab{U(y) - U(z)}}{\kappa D}}.
\end{gather}
As Eq.~\eqref{eq:T.infty} indicates, greater degree heterogeneity increases the effective noise intensity, thereby reducing mean escape times.
Importantly, no other characteristics of the underlying network affect the collective escape dynamics in this limit, underscoring the significance of $\kappa$.

\section{Effective one-dimensional dynamics}
Extending our analysis of the fully connected model~\jointarticle{}, we derive effective one-dimensional dynamics that describe collective processes of noise-induced escape.
Our primary concern is the effect of varying degree heterogeneity $\kappa$, in addition to coupling strength $K$.

\subsection{Nonlinear mean-field Fokker--Planck equation}
First, we derive a one-dimensional reduced description called nonlinear mean-field Fokker--Planck equation (NlinMFFPE) assuming weak coupling.
The evolution of the $N$-variate probability density function (PDF) $p_N(x_1, \ldots, x_N, t)$ is governed by the Fokker--Planck equation (FPE) associated with the model [Eq.~\eqref{eq:approx.model}]:
\begin{gather} \label{eq:pN.FPE} \begin{split}
  \partial_t p_N =& -\sum_{j=1}^N \partial_{x_j} \Bab{
    \bab{f(x_j) + K \sum_{m=1}^N \rho_m \pab{x_m - x_j}} p_N
  }  \\
  & + D \sum_{i,j} \partial_{x_i x_j} p_N.
\end{split} \end{gather}
To reduce the degrees of freedom, we marginalize $p_N$ with respect to $x_2, \ldots, x_N$ to obtain the univariate PDF of $x_1$, i.e.,
\begin{gather}
  p_{1(1)}(x_1, t) \coloneq \int \prod_{l=2}^N \d x_l \, p_N(x_1, \ldots, x_N, t).
\end{gather}
By marginalizing the FPE~\eqref{eq:pN.FPE} and dropping surface terms, the dynamical equation for $p_{1(1)}$ becomes
\begin{gather} \label{eq:bbgky} \begin{split}
  \partial_t p_{1(1)} 
    =& -\partial_{x_1} \bigggl\{
    \bab{f(x_1) - K \pab{1 - \rho_1} x_1} p_{1(1)}  \\
    & + K \sum_{m=2}^N \rho_m \int \d x_m x_m p_{2(1,m)}(x_1, x_m, t)
  \bigggr\}  \\
    & + D \partial_{x_1}^2 p_{1(1)},
\end{split} \end{gather}
where $p_{2(i,j)}(x_i, x_j, t)$ denotes the bivariate PDF for $x_i$ and $x_j$.
Eq.~\eqref{eq:bbgky} shows that the evolution of the univariate PDF $p_{1(1)}$ depends on the bivariate PDFs $p_{2(1,m)}$.
Such hierarchical dependence among PDFs is known as the BBGKY hierarchy.

To close the hierarchy, we first assume independence between nodes, i.e.,
\begin{gather}
  p_{2(i,j)}(x_i, x_j, t) \approx p_{1(i)}(x_i, t) p_{1(j)}(x_j, t),
\end{gather}
which should be valid for sufficiently weak coupling (i.e., for small $K$).
This assumption simplifies the expectation of $\Theta$ as follows:
\begin{subequations} \begin{align}
  \aab{\Theta} =& \int \d x_1 \dots \d x_N \sum_{j=1}^N \rho_j x_j p_N(x_1, \dots, x_N, t)  \\
    \approx& \sum_{j=1}^N \rho_j \int \d x_j \, x_j p_{1(j)}(x_j, t)  \\
    =& \sum_{j=1}^N \rho_j \aab{x_j}_j,
\end{align} \end{subequations}
where $\aab{x_j}_j$ denotes the expectation with respect to $p_{1(j)}$.
Accordingly, Eq.~\eqref{eq:bbgky} becomes
\begin{gather} \label{eq:fpe.rhoi} \begin{split}
  \partial_t p_{1(i)}
  =& -\partial_{x_i} \biggl\{
    \biggl[ f(x_i) + K \biggl(\aab{\Theta} - x_i -  \\
  & \qquad \rho_i \pab{ \aab{x_i}_{i} - x_i } \biggr) \biggr] p_{1(i)}
  \biggr\} + D \partial_{x_i}^2 p_{1(i)}.
\end{split} \end{gather}
For nodes with negligible $\rho_i$ satisfying $\rho_i / N \to 0$ as $N \to \infty$, the evolution equation of $p_{1(i)}$ [Eq.~\eqref{eq:fpe.rhoi}] no longer depends on the node index $i$.
Considering that the initial condition of $p_{1(i)}$ is the same across nodes in our setting, the PDFs $p_{1(i)}$ for nodes with negligible $\rho_i$ would be approximately identical.
Hence, when $\max_i \rho_i$ vanishes as $N \to \infty$, all nodes would be statistically indistinguishable.
That is, the univariate PDF would be the same across nodes:
\begin{gather}
  p_{1(i)}(x_i, t) \approx p_1(x, t).
\end{gather}
It follows that $\aab{x_j}_j \approx \aab{x}$, where $\aab{x}$ denotes the expectation of $p_1$, which implies $\aab{\Theta} \approx \aab{x}$.
In the degree-homogeneous limit of $\kappa = 1$, it holds that $\rho_i = 1 / N$ for all $i$, and hence $\rho_i \to 0$ as $N \to \infty$.
Taken together, when $K$ and $\kappa$ are sufficiently small, we obtain the following NlinMFFPE for large $N$:
\begin{gather} \label{eq:nlinmffpe}
  \partial_t p_1 = 
    -\partial_x \Bab{\bab{f(x) + K \pab{\aab{x} - x}} p_1}  
    + D \partial_x^2 p_1.
\end{gather}
While this is a closed system for $p_1(x, t)$, it is nonlinear in $p_1$ because the expectation $\aab{x}$ appears in the drift term.
In the literature, the procedure discussed above is known as the Weiss mean-field approximation~\cite{vandenbroeck1994NoiseInduced}.
Importantly, NlinMFFPE suggests that properties of the underlying network may be irrelevant to the evolution of $p_1$ for sufficiently weak coupling.
Indeed, it is exactly the same as the NlinMFFPE derived from the fully connected model in the companion Letter~\jointarticle{}.
Accordingly, the same procedure based on the probability current allows us to predict mean escape times.

\subsection{Stochastic mean-field dynamics}
Next, we present a networked formulation of stochastic mean-field dynamics (SMFD), which is valid under strong coupling.
The derivation is inspired by a previous study on a model of elastically coupled hair bundles in the strong-coupling limit~\cite{dierkes2012Meanfield}.
Here, we reformulate the approach to make it applicable to our model at finite coupling strength.
A central quantity in SMFD is the weighted mean square deviation $Z(t)$ defined in Eq.~\eqref{eq:Z}.
The dynamics of $\Theta$ [Eq.~\eqref{eq:theta.dyn}] involve $Z$ and the effective noise $\hat{\eta}_\Theta$.
We reduce the degrees of freedom of the $\Theta$ dynamics by introducing auxiliary one-dimensional stochastic processes that approximate $Z$ and $\hat{\eta}_\Theta$.
By differentiating $Z$ with respect to time and using Eq.~\eqref{eq:yi.dyn}, one obtains
\begin{gather} \label{eq:Z.sde}
  \frac{\dot{Z}}{2} = -\bab{K - f'(\Theta)} Z + \sqrt{2D} \hat{\eta}_Z,
\end{gather}
where $\hat{\eta}_Z$ is the effective noise defined as
\begin{gather} \label{eq:etaZ.def}
  \hat{\eta}_Z(t) \coloneq \sum_{j=1}^N \rho_j y_j(t) \hat{\eta}_j(t).
\end{gather}
The product $y_j(t) \hat{\eta}_j(t)$ is understood in the Ito sense.
To obtain a closed evolution equation for $Z$, we approximate $\hat{\eta}_Z$ by a single effective noise source while preserving its statistical properties.
More precisely, we calculate its first and second moments and its correlation function.
The first step is to write down the formal solution for $y_i$, which approximately obeys the OU process [Eq.~\eqref{eq:yi.dyn}].
Then, substituting the solution into Eq.~\eqref{eq:etaZ.def} allows us to evaluate its moments under a technical assumption.
Details are provided in Appendix~\ref{app:smfd}.
For conciseness, we neglect fluctuations in $\hat{\eta}_Z$ and retain only its mean contribution in the main text.
The mean of $\hat{\eta}_Z(t)$ is given by
\begin{gather}
  \aab{\hat{\eta}_Z}(t) = \sqrt{\frac{D}{2}} \pab{1 - \frac{\kappa}{N}}.
\end{gather}
Replacing $\hat{\eta}_Z(t)$ in Eq.~\eqref{eq:Z.sde} with its mean, we obtain the following system:
\begin{subequations} \label{eq:smfd.2d} \begin{align}
  \dot{\Theta} =& f(\Theta) + \frac{f''(\Theta)}{2} Z + \sqrt{2D} \sqrt{\frac{\kappa}{N}} \eta_\Theta,  \\
  \frac{\dot{Z}}{2} =& -\bab{K - f'(\Theta)} Z + D \pab{1 - \frac{\kappa}{N}}.
\end{align} \end{subequations}

The derived system [Eq.~\eqref{eq:smfd.2d}] indicates that, for sufficiently large $K$ satisfying $K \gg f(\Theta)$ and $K \gg f'(\Theta)$, $Z$ evolves much faster than $\Theta$.
Therefore, one may further simplify the system through adiabatic elimination of $Z$, yielding the following one-dimensional SDE for $\Theta$:
\begin{subequations} \label{eq:smfd}
\begin{align} \label{eq:smfd.theta}
  & \dot{\Theta} = f(\Theta) + \frac{f''(\Theta)}{2} Z^*(\Theta) + \sqrt{2D} \sqrt{\frac{\kappa}{N}} \eta_\Theta,  \\
  \label{eq:smfd.zstar}
  & Z^*(\Theta) \coloneq \frac{D}{K - f'(\Theta)} \pab{1 - \frac{\kappa}{N}}.
\end{align}
\end{subequations}
This is the one-dimensional SMFD for networked systems.
As it has an associated potential, one can calculate mean escape times using the formula [Eq.~\eqref{eq:mfpt}].
Details are provided in Appendix~\ref{app:smfd}.

SMFD clearly delineate the interplay of nonlinearity, coupling, and noise in the collective escape process.
The second term in the $\Theta$ dynamics involves both coupling strength $K$ and noise intensity $D$ through $Z^*$.
Noise induces a finite diversity $Z^*$ within the network, which systematically drives the mean field $\Theta$ due to the system's nonlinearity, reflected in the nonvanishing $f''(\Theta)$.
Eq.~\eqref{eq:smfd.zstar} indicates that greater degree heterogeneity (i.e., larger $\kappa$) shrinks diversity $Z^*$, thereby weakening its facilitation of escape.
At the same time, larger $\kappa$ increases the effective noise intensity $D \kappa / N$, accelerating escape.
To conclude, we note that the networked SMFD [Eq.~\eqref{eq:smfd}] generalize the SMFD derived for the fully connected model~\jointarticle{}.
Indeed, by setting $\rho_i \equiv 1 / N$ and $\kappa = 1$, Eq.~\eqref{eq:smfd} coincides with the SMFD presented in the companion Letter~\jointarticle{} for large $N$.
In the global coupling case, correlations between $y_i$ and $y_j$ ($i \neq j$) are negligible for large $N$, as seen from the correlation function [Eq.~\eqref{eq:corr.etahati}].
As such, the derivation of SMFD is much simpler in fully connected cases.

In the companion Letter~\jointarticle{}, we determined the boundary between weak-coupling and intermediate regimes, $K_1$, based on a necessary condition for the validity of SMFD.
Since deriving SMFD relies on the OU approximation of the $y_i$ dynamics [Eq.~\eqref{eq:yi.dyn}], the distribution of $y_i$ (or equivalently $x_i$) should be approximately Gaussian in the regime where SMFD apply.
We neglected the third-order term involving $m_3 \coloneq \sum_{j=1}^N \rho_j y_j^3$ in obtaining the $\Theta$ dynamics [Eq.~\eqref{eq:theta.dyn}].
In degree-homogeneous cases where $\kappa \approx 1$, $m_3$ corresponds to the third moment of $y_i$, which vanishes if $y_i$ follows a Gaussian distribution.
Accordingly, we could estimate $K_1$ based on the dynamical properties of the fully connected model.
In contrast, heterogeneity in $\rho_j$ prevents $m_3$ from vanishing in general.
Qualitatively, one enters the intermediate regime once $K$ becomes sufficiently large that $m_3$ can be neglected.
As $m_3$ depends on $\rho_j$, the boundary $K_1$ should be a function of $\kappa$.
A quantitative estimation of $K_1$ is left for future study.

\subsection{Deterministic mean-field dynamics}

\begin{figure*}[t]
  \centering
  \includegraphics[width=\linewidth]{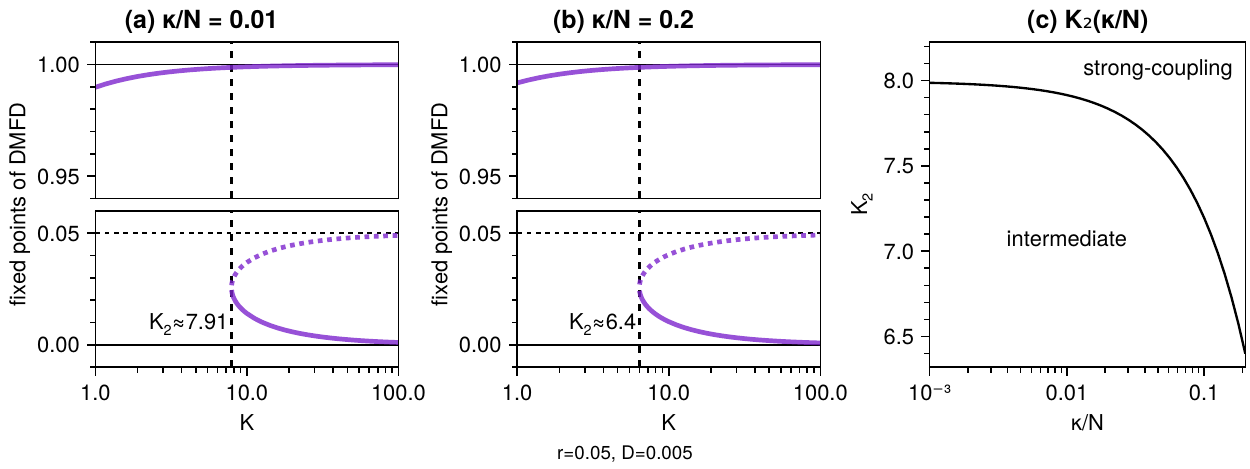}
  \caption{%
    The bifurcation point of DMFD, $K_2$, depends on the normalized degree heterogeneity $\kappa / N$.
    (a) Bifurcation diagram for $\kappa / N = 0.01$.
    (b) Bifurcation diagram for $\kappa / N = 0.2$.
    (c) The bifurcation point $K_2$ as a function of $\kappa / N$.
  }
  \label{fig:dmfd}
\end{figure*}

When the time scale of the diffusion is much slower than that of the drift in SMFD, one may ignore the noise term to obtain the following deterministic mean-field dynamics (DMFD):
\begin{gather} \label{eq:dmfd}
  \dot{\Theta} = g\pab{\Theta; K, \frac{\kappa}{N}}
    \coloneq f(\Theta) + \frac{f''(\Theta)}{2} \frac{D}{K} \pab{1 - \frac{\kappa}{N}}.
\end{gather}
Here, we approximated $K - f'(\Theta) \approx K$ for large $K$.
As shown in Fig.~\ref{fig:dmfd}(a) and (b), DMFD exhibit a saddle-node bifurcation, where the system changes from monostable to bistable.
In the monostable regime where $K$ is smaller than the bifurcation point $K_2$, the collective active state $\Theta \approx 1$ is the only stable state of the system.
This indicates that the diversity within the system, $Z^*$, induces escape even without fluctuation in $\Theta$.
Put differently, noise-induced escape is impossible in the limit $\kappa / N \to 0$ in the bistable regime beyond $K_2$.

The bifurcation point, which depends on $\kappa / N$, is where the minimum of $g(\Theta; K, \kappa / N)$ equals zero.
Solving $\d g / \d \Theta = 0$, one finds that the location of the minimum is
\begin{gather}
  \Theta_0 = \frac{1}{3} \bab{1 + r - \sqrt{1 - r + r^2 - \frac{9D}{K} \pab{1 - \frac{\kappa}{N}}}}.
\end{gather}
By solving $g(\Theta_0; K, \kappa / N) = 0$ for $K$ with different values of $\kappa / N$, one numerically obtains $K_2(\kappa / N)$, plotted in Fig.~\ref{fig:dmfd}(c).
We distinguish the intermediate and strong-coupling regimes at $K_2$.

When all nodes have the same degree (i.e., $\kappa = 1$), DMFD can be derived from NlinMFFPE in the limit $N \to \infty$, as discussed in the companion Letter~\jointarticle{}.
In such cases, one may regard DMFD as describing the evolution of the expectation with respect to $p_1(x, t)$ [Eq.~\eqref{eq:nlinmffpe}].
In degree-heterogeneous cases, a similar relation between NlinMFFPE and DMFD is expected; however, it is not straightforward to derive DMFD [Eq.~\eqref{eq:dmfd}] from NlinMFFPE [Eq.~\eqref{eq:nlinmffpe}] due to the (out)degree weighting.

\subsection{Summary of theoretical insights}
So far, we first simplified the model through the annealed network approximation and then derived three types of effective dynamics.
The theoretical analyses offered five major insights.
(i) The results from the study of fully connected models~\jointarticle{} remain qualitatively valid in degree-heterogeneous cases.
That is, the interplay of nonlinearity, coupling, and noise yields three qualitatively distinct escape mechanisms, which are represented by NlinMFFPE, DMFD, and SMFD.
(ii) When coupling is sufficiently weak, NlinMFFPE allows us to predict mean escape times.
Notably, this implies that properties of the underlying network, including degree heterogeneity $\kappa$, would not affect escape dynamics for sufficiently small $K$.
(iii) Systems on highly heterogeneous networks (i.e., networks with large $\kappa / N$) are the first to deviate from predictions of NlinMFFPE.
(iv) For sufficiently large $K$, the networked SMFD [Eq.~\eqref{eq:smfd}] allow us to predict mean escape times as a function of $K$ and $\kappa / N$.
(v) As $K$ increases, mean escape times increase and approach asymptotic values $T_\infty(\kappa/N)$ [Eq.~\eqref{eq:T.infty}], which depend on (out)degree heterogeneity.
Put differently, network characteristics other than $\kappa / N$ do not affect escape dynamics under sufficiently strong coupling.

Taken together, these results show that degree heterogeneity shapes not only the quantitative values of mean escape times but also the range of coupling strengths over which each reduced description provides an accurate account of collective escape.

\section{Numerical validation}
To validate our theoretical insights, we performed direct numerical simulations of the original model [Eq.~\eqref{eq:original.model}].
For multiple values of $K$, we numerically measured mean escape times while varying the normalized degree heterogeneity $\kappa / N$.
In the following, we restrict our attention to undirected networks.
We also focus on NlinMFFPE and SMFD because DMFD are expected to be valid only over a limited range of $K$.
In addition, although DMFD offer an interesting perspective on the mechanism of collective escape, SMFD suffice for estimating mean escape times.

\begin{figure}[t]
  \centering
  \includegraphics[width=\linewidth]{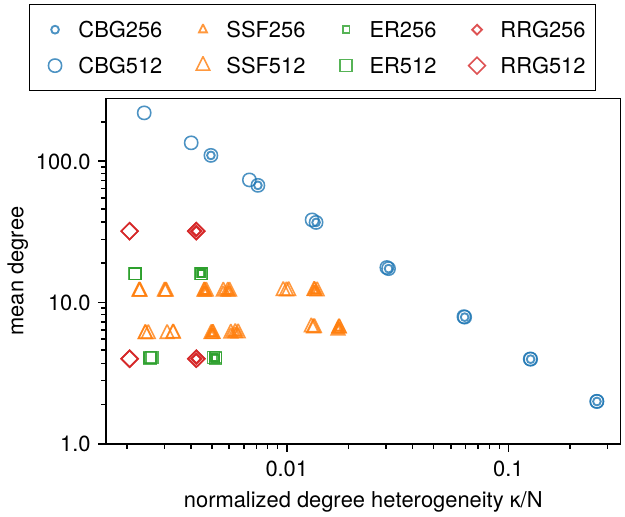}
  \caption{%
    Normalized degree heterogeneity $\kappa / N$ and mean degree of networks used in the numerical analysis.
  }
  \label{fig:meandeg.vs.deghet}
\end{figure}

To cover a range of networks from homogeneous to highly heterogeneous, we used four network models to generate network instances.
Fig.~\ref{fig:meandeg.vs.deghet} shows the mean degrees and normalized degree heterogeneities of the generated networks.
%
The first network model is the complete bipartite graph (CBG), where nodes are partitioned into two subsets of sizes $n_1$ and $n_2 = N - n_1$.
Each node is connected to every node in the other subset.
Degree heterogeneity $\kappa$ of a CBG is calculated as
\begin{gather}
  \kappa = \frac{\aab{d^2}}{\aab{d}^2} = \frac{N^2}{4 n_1 n_2}.
\end{gather}
With $n_1 = n_2 = N / 2$, the resulting network is homogeneous with $\kappa = 1$.
In the other limit, $n_1 = N - 1$ and $n_2 = 1$, the resulting network is a star, which is highly heterogeneous with $\kappa = \order{N}$.
%
The second network model is the random regular graph (RRG), in which all nodes have the same degree.
Accordingly, it always holds that $\kappa = 1$ for a RRG.
We used RRGs to obtain homogeneous networks.
%
The third model is the static scale-free (SSF) network~\cite{goh2001Universal}.
In the limit $N \to \infty$, the second moment $\aab{d^2}$, and hence $\kappa$, of a scale-free network diverges for $\alpha \leq 3$~\cite{newman201810}, where $\alpha$ is the exponent of the distribution (i.e., $p(d) \sim d^{-\alpha}$).
We used SSF networks with different exponents $\alpha \geq 2$ as examples of highly heterogeneous networks.
We note, however, that the realized $\kappa$ values of the generated networks (Fig.~\ref{fig:meandeg.vs.deghet}) were not large ($\kappa / N \sim 0.01$).
This was probably because the system size was too small ($N \sim 100$) to produce nodes with extremely large degrees.
%
The last model is the Erd\H{o}s--R\'{e}nyi (ER) network, whose degree distribution approaches a Poisson distribution in the limit of large $N$~\cite{newman201811}.
That is, $p(d) \approx e^{-c} c^d / d!$, where $c = \aab{d}$ is the mean degree.
Accordingly, degree heterogeneity is $\kappa = 1 + 1 / c$. 
This implies that $\kappa / N$ would be small for large $N$: i.e., an ER network is nearly degree-homogeneous.
Numerical methods are detailed in Appendix~\ref{app:num.methods}.

\begin{figure*}[t]
  \centering
  \includegraphics[width=\linewidth]{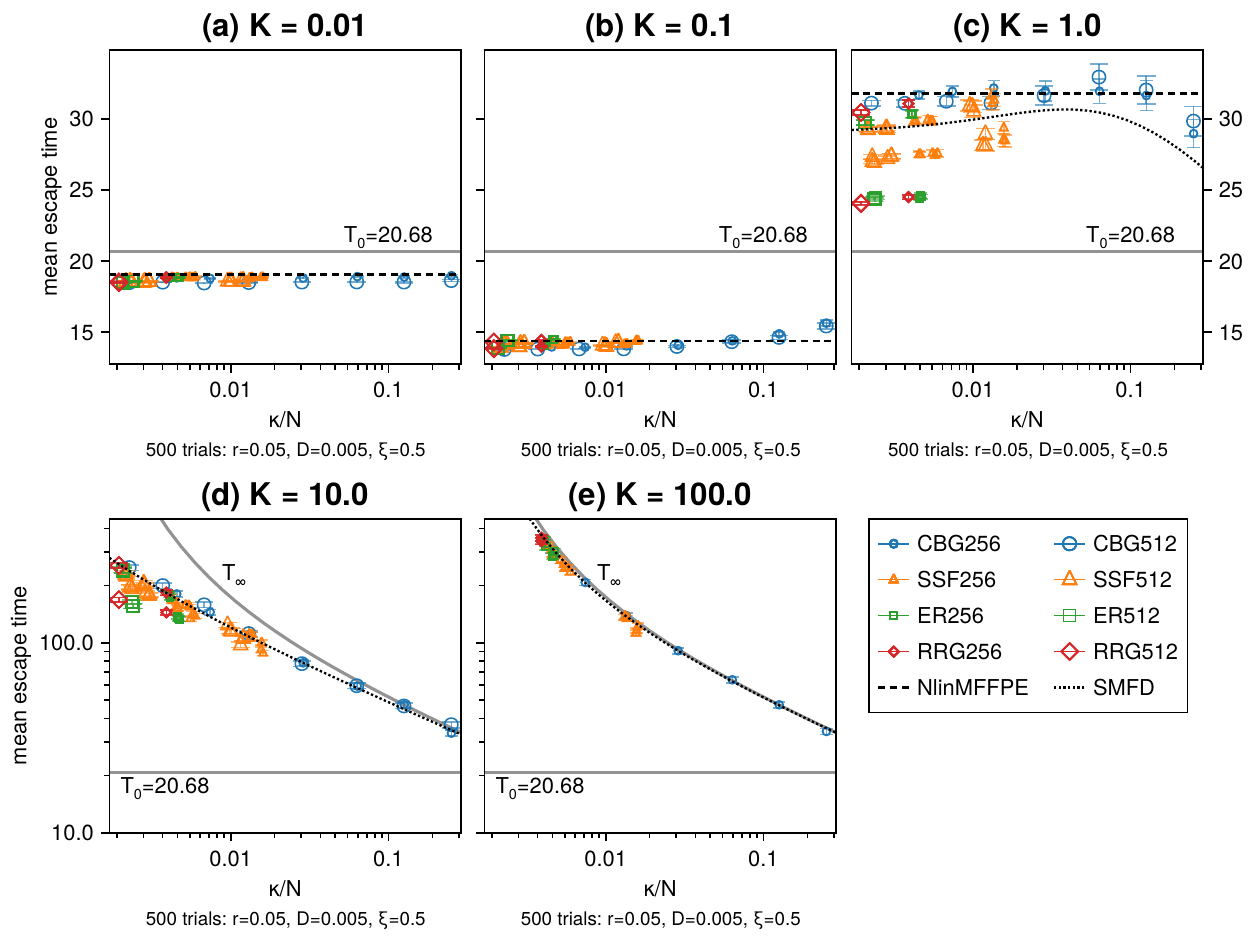}
  \caption{%
    Numerically measured mean escape times (markers) and theoretical predictions (lines) as functions of normalized degree heterogeneity $\kappa / N$ at different values of coupling strength $K$.
    The mean escape time of the uncoupled system (i.e., $K = 0$) is denoted by $T_0$ [Eq.~\eqref{eq:T0}].
    Marker shape and size indicate the network model and the number of nodes, respectively.
    Error bars represent standard errors.
    (a) The influence of $\kappa / N$ was not visible at $K = 0.01$.
    (b) Systems on networks with large $\kappa / N$ deviated from the prediction of NlinMFFPE at $K = 0.1$.
    (c) Network characteristics, including $\kappa / N$, strongly affected mean escape times at $K = 1$.
    (d) Deviations from the prediction of SMFD shrank as $K$ increased.
    (e) At $K = 100$, the numerical results and the prediction of SMFD agreed well and coincided with $T_\infty(\kappa / N)$ [Eq.~\eqref{eq:T.infty}].
  }
  \label{fig:met.vs.deghet}
\end{figure*}

\begin{figure*}[t]
  \centering
  \includegraphics[width=\linewidth]{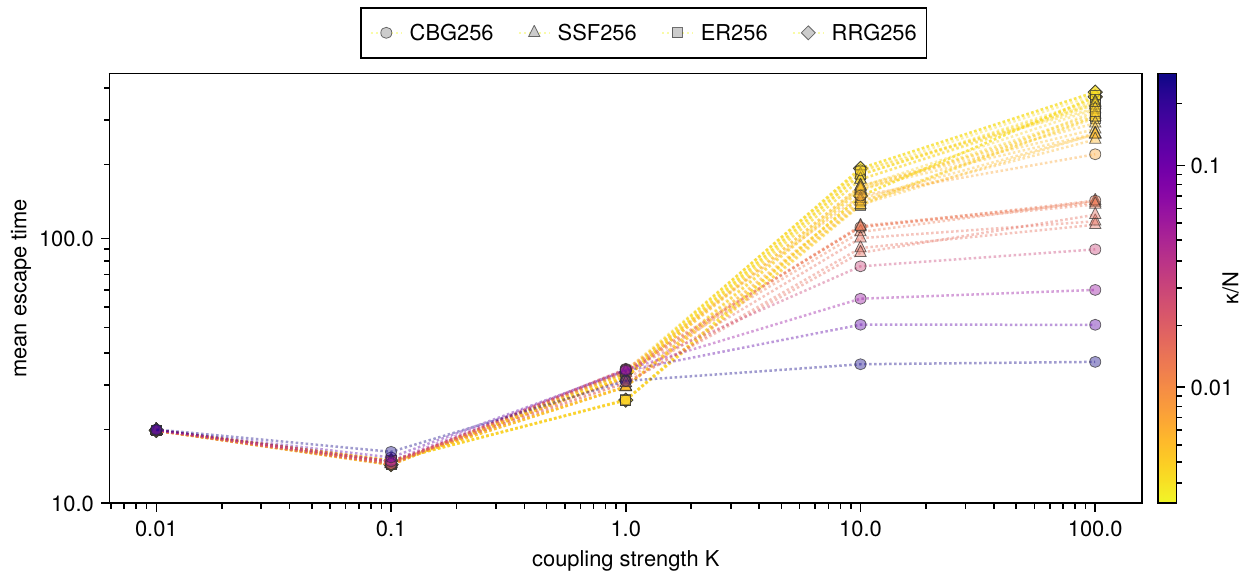}
  \caption{%
    Dependence of the numerically measured mean escape times on the coupling strength $K$.
    Larger values of $\kappa / N$ (color-coded) suppressed the influence of $K$.
  }
  \label{fig:met.vs.K}
\end{figure*}

The numerical results are summarized in Fig.~\ref{fig:met.vs.deghet}.
Overall, they supported the theoretical insights listed above.
At $K = 0.01$ [panel~(a)], mean escape times declined from the reference value of the uncoupled system, $T_0$ [Eq.~\eqref{eq:T0}].
Degree heterogeneity $\kappa / N$ does not appear to have affected mean escape times.
This suggests that, although we derived NlinMFFPE assuming small $\kappa$, even star graphs were not heterogeneous enough to invalidate NlinMFFPE at $K = 0.01$.
At $K = 0.1$ [panel~(b)], networks with the highest degree heterogeneity deviated slightly from the prediction of NlinMFFPE.
At the intermediate coupling strength $K = 1$ [panel~(c)], while both NlinMFFPE and SMFD captured the numerical results qualitatively, differences across network instances were visible even among those with similar $\kappa / N$ values.
As $K$ increased to $K = 10$ [panel~(d)], simulation results clustered around the theoretical curve based on SMFD.
Deviations from the prediction of SMFD became even smaller at $K = 100$ [panel~(e)].
In addition, panel~(e) indicates that mean escape times mostly reached the limiting values, $T_\infty$ [Eq.~\eqref{eq:T.infty}], at $K = 100$.

According to Fig.~\ref{fig:met.vs.deghet}, the case $K = 1$ appears to be marginal, in the sense that neither NlinMFFPE nor SMFD predicts mean escape times accurately.
On the one hand, the interaction is too strong to ignore the underlying network structure, invalidating NlinMFFPE.
On the other hand, it is too weak to characterize each element solely by its degree, invalidating the annealed network approximation in the first place.
That is, other properties, such as the mean degree of neighbors, are expected to affect the evolution of each element's state.
Another source of deviations from the predictions of SMFD would be the difference in the definitions of mean escape times.
While the within-system average of escape times was measured in the simulations, predictions of SMFD represent the expected escape time of the mean field $\Theta$.
Although these two mean escape times coincide under sufficiently strong coupling, the escape of the mean field is expected to occur earlier than the within-system average, which is indeed the case in Fig.~\ref{fig:met.vs.deghet}(c).
We note that the prediction of SMFD for $K = 1$ is nonmonotonic in $\kappa / N$.
This is because, as seen from SMFD [Eq.~\eqref{eq:smfd.theta}], larger degree heterogeneity both facilitates and impedes escape.
While larger $\kappa$ results in stronger effective noise and accelerates escape, it also reduces the diversity $Z^*(\Theta)$, impeding escape.

As illustrated in Fig.~\ref{fig:met.vs.K}, the numerical results indicate that the qualitative dependence of mean escape times on coupling strength $K$ is similar across degree heterogeneities.
Indeed, mean escape times were shorter than $T_0$ at $K = 0.01$ and even shorter at $K = 0.1$.
Then they increased as the coupling became stronger, approaching $T_\infty(\kappa / N)$.
In addition, the results suggest that a high degree heterogeneity suppresses the influence of coupling strength on the escape dynamics.
For instance, the decline in mean escape times from $T_0$ at $K = 0.1$ was smaller for networks with higher degree heterogeneity.
Similarly, the subsequent increase in mean escape times was significantly smaller for more heterogeneous networks.

\section{Discussion}
This article studied noise-induced escape in a class of networked populations of bistable elements.
Through the annealed network approximation, we obtained the approximate model [Eq.~\eqref{eq:approx.model}] that could be expressed in terms of the degree-weighted mean field $\Theta$ and deviations $y_i$.
We then derived three reduced descriptions --- nonlinear mean-field Fokker--Planck equation (NlinMFFPE) [Eq.~\eqref{eq:nlinmffpe}], deterministic mean-field dynamics (DMFD) [Eq.~\eqref{eq:dmfd}], and stochastic mean-field dynamics (SMFD) [Eq.~\eqref{eq:smfd}].
Each reduced description represents a qualitatively distinct mechanism of collective escape.
We validated theoretical predictions for mean escape times by direct numerical simulations of the original model.
Our analysis revealed the dominant roles of coupling strength $K$ and degree heterogeneity $\kappa$, which quantifies variability in (out)degree.
That is, the validity and dynamical properties of the reduced descriptions depend on both coupling strength and degree heterogeneity.
The companion Letter~\jointarticle{} studied the influence of varying coupling strength in the fully connected setting.
This article showed that degree heterogeneity, in addition to coupling strength, shapes escape mechanisms in networked bistable elements.

Assuming weak coupling, we derived NlinMFFPE [Eq.~\eqref{eq:nlinmffpe}], which enabled predictions of mean escape times.
However, its derivation required the assumption of low degree heterogeneity in addition to weak coupling.
For strong coupling, we derived the networked SMFD [Eq.~\eqref{eq:smfd}], which represent effective one-dimensional dynamics of $\Theta$.
SMFD depend both on coupling strength $K$ and on normalized degree heterogeneity $\kappa/N$.
Accordingly, the estimate of mean escape times based on SMFD is determined by $K$ and $\kappa/N$.
Direct numerical simulations of the original model [Eq.~\eqref{eq:original.model}] on various networks supported these insights.
In particular, while the dependence of mean escape times on $K$ was qualitatively similar across degree heterogeneities, greater heterogeneity attenuated the effect of varying $K$.
This is likely due to the strong correlation between influential nodes and the degree-weighted mean field $\Theta$.
In our model [Eq.~\eqref{eq:original.model}], higher degree heterogeneity undermines the influence of low-degree nodes on the mean field, weakening the effective feedback mediated by diffusive coupling.

Our results suggest the following relation between degree heterogeneity and escape dynamics.
When the coupling is too weak, properties of the underlying network do not affect the dynamics.
Put differently, all that matters is whether --- not how --- a node interacts with others.
With slightly stronger coupling, the effects of degree heterogeneity and other network characteristics become apparent.
In this regime, the coupling is too strong to neglect the contribution of network structure but too weak to regard the system as effectively one-dimensional.
As one increases the coupling strength further, all nodes come to follow the mean field.
Since the effective mean-field dynamics depend only on degree heterogeneity, and not on other network properties, degree heterogeneity alone becomes sufficient for characterizing collective escape.

We briefly discuss the scope and limitations of the present results, as well as several directions for future work.
First, we expect that our results are insensitive to the definition of mean escape times.
Nevertheless, in highly heterogeneous networks, the escape of a hub node may exhibit distinct behavior, because the weighted mean field $\Theta$ is strongly correlated with high-degree nodes.
Second, our results crucially depend on the diffusive nature of the coupling term.
Thus, it would be challenging to extend our analysis to non-diffusive couplings.
Third, significant heterogeneity in local dynamics --- such as in $r$ and $D$ --- is beyond the scope of the present analysis.
Finally, the coupling term of the original model is normalized by the focal node's indegree $\rmsup{d}{in}_i$.
This normalization was essential for deriving the simplified approximate model [Eq.~\eqref{eq:approx.model}] in terms of $\Theta$ and $y_i$.
Relaxing this modeling choice, and identifying other settings that yield a similar $\rho_j$-based reduction, would be an interesting direction for future research.

We conclude by highlighting that the SMFD approach provides a useful perspective on the interplay of nonlinearity, diffusive coupling, and dynamical noise.
The SMFD formulation is inspired by Dierkes and colleagues~\cite{dierkes2012Meanfield}, who focused on the strong-coupling limit, whereas our formulation refines the approach to obtain an effective one-dimensional description at finite coupling strength.
The obtained SMFD concisely illustrate how dynamical noise generates both the effective fluctuation of the mean field and the diversity within the network, and how these noise-induced effects, together with the nonlinearity of the local dynamics and the diffusive coupling, synergistically drive collective escape.
In principle, we expect similar phenomena in diffusively coupled random dynamical systems, and the present SMFD approach should be applicable to such situations.

\begin{acknowledgments}
H.I. acknowledges support from the World-leading Innovative Graduate Study Program in Proactive Environmental Studies (WINGS-PES), the University of Tokyo, and JSPS KAKENHI Grant Number JP24KJ0635.
This study was also supported by the JSPS Core-to-Core Program ``Advanced core-to-core network for the physics of self-organizing active matter'' JPJSCCA20230002 to H.I. and H.K.
\end{acknowledgments}

\section*{Data Availability}
The data and code that support the findings of this study are openly available at \url{https://github.com/ishiihidemasa/26-escape-network}.

\appendix

\begin{table}[tpb]
  \centering
  \caption{%
    Parameters and degree heterogeneity $\kappa$ of the complete bipartite graphs (CBGs) used in our numerical analysis.
  }
  \label{tab.deghet:cbg.param}
  \begin{tabular}{lrrr@{\hspace{12mm}}lrrr}
    \toprule
    id & $N$ & $n_2$ & $\kappa$ & id & $N$ & $n_2$ & $\kappa$  \\
    \midrule
    cbg256-1 & 256 &  1 & 64.3 & cbg512-1 & 512 &   1 &  128 \\
    cbg256-2 & 256 &  2 & 32.3 & cbg512-2 & 512 &   2 & 64.3 \\
    cbg256-3 & 256 &  4 & 16.3 & cbg512-3 & 512 &   4 & 32.3 \\
    cbg256-4 & 256 &  9 & 7.37 & cbg512-4 & 512 &   9 & 14.5 \\
    cbg256-5 & 256 & 20 & 3.47 & cbg512-5 & 512 &  20 & 6.66 \\
    cbg256-6 & 256 & 40 & 1.90 & cbg512-6 & 512 &  40 & 3.47 \\
    cbg256-7 & 256 & 80 & 1.16 & cbg512-7 & 512 &  80 & 1.90 \\
             &     &    &      & cbg512-8 & 512 & 160 & 1.16 \\
    \bottomrule
  \end{tabular}
\end{table}

\begin{table}[tpb]
  \centering
  \caption{%
    Parameters and degree heterogeneity $\kappa$ of the static scale-free networks (SSFs) used in our numerical analysis.
  }
  \label{tab.deghet:ssf.param}
  \begin{tabular}{lrrrrr}
    \toprule
    id & $N$ & $m$ & $\gamma$ & seed & $\kappa$  \\
    \midrule
     ssf256-1 & 256 & 3 & 2.0 & 1 & 4.11  \\ 
     ssf256-2 & 256 & 3 & 2.0 & 2 & 4.02  \\
     ssf256-3 & 256 & 3 & 2.0 & 3 & 3.99  \\
     ssf256-4 & 256 & 3 & 2.4 & 1 & 1.17  \\
     ssf256-5 & 256 & 3 & 2.4 & 2 & 1.18  \\
     ssf256-6 & 256 & 3 & 2.4 & 3 & 1.19  \\
     ssf256-7 & 256 & 3 & 3.4 & 1 & 1.42  \\
     ssf256-8 & 256 & 3 & 3.4 & 2 & 1.47  \\
     ssf256-9 & 256 & 3 & 3.4 & 3 & 1.53  \\
    ssf256-10 & 256 & 6 & 2.0 & 1 & 3.40  \\
    ssf256-11 & 256 & 6 & 2.0 & 2 & 3.50  \\
    ssf256-12 & 256 & 6 & 2.0 & 3 & 3.36  \\
    ssf256-13 & 256 & 6 & 2.4 & 1 & 1.08  \\
    ssf256-14 & 256 & 6 & 2.4 & 2 & 1.09  \\
    ssf256-15 & 256 & 6 & 2.4 & 3 & 1.11  \\
    ssf256-16 & 256 & 6 & 3.4 & 1 & 1.32  \\
    ssf256-17 & 256 & 6 & 3.4 & 2 & 1.39  \\
    ssf256-18 & 256 & 6 & 3.4 & 3 & 1.41  \\
    \midrule
    id & $N$ & $m$ & $\gamma$ & seed & $\kappa$  \\
    \midrule
     ssf512-1 & 512 & 3 & 2.0 & 1 & 6.14  \\
     ssf512-2 & 512 & 3 & 2.0 & 2 & 6.16  \\  
     ssf512-3 & 512 & 3 & 2.0 & 3 & 5.92  \\  
     ssf512-4 & 512 & 3 & 2.4 & 1 & 1.18  \\  
     ssf512-5 & 512 & 3 & 2.4 & 2 & 1.18  \\  
     ssf512-6 & 512 & 3 & 2.4 & 3 & 1.21  \\  
     ssf512-7 & 512 & 3 & 3.4 & 1 & 1.47  \\  
     ssf512-8 & 512 & 3 & 3.4 & 2 & 1.55  \\  
     ssf512-9 & 512 & 3 & 3.4 & 3 & 1.56  \\  
    ssf512-10 & 512 & 6 & 2.0 & 1 & 4.88  \\  
    ssf512-11 & 512 & 6 & 2.0 & 2 & 5.18  \\  
    ssf512-12 & 512 & 6 & 2.0 & 3 & 5.07  \\  
    ssf512-13 & 512 & 6 & 2.4 & 1 & 1.10  \\  
    ssf512-14 & 512 & 6 & 2.4 & 2 & 1.11  \\  
    ssf512-15 & 512 & 6 & 2.4 & 3 & 1.11  \\  
    ssf512-16 & 512 & 6 & 3.4 & 1 & 1.44  \\  
    ssf512-17 & 512 & 6 & 3.4 & 2 & 1.46  \\  
    ssf512-18 & 512 & 6 & 3.4 & 3 & 1.45  \\  
    \bottomrule
  \end{tabular}
\end{table}

\begin{table}[tpb]
  \centering
  \caption{%
    Parameters and degree heterogeneity $\kappa$ of the Erdős--Rényi networks (ERs) used in our numerical analysis.
  }
  \label{tab.deghet:er.param}
  \begin{tabular}{lrrrr@{\hspace{12mm}}lrrrr}
    \toprule
    id & $N$ & $m$ & seed & $\kappa$ & id & $N$ & $m$ & seed & $\kappa$  \\
    \midrule
    er256-1 & 256 &  4 & 1 & 1.17 & er512-1 & 512 &  4 & 1 & 1.21 \\
    er256-2 & 256 &  4 & 2 & 1.19 & er512-2 & 512 &  4 & 2 & 1.23 \\
    er256-3 & 256 &  4 & 3 & 1.21 & er512-3 & 512 &  4 & 3 & 1.23 \\
    er256-4 & 256 & 16 & 1 & 1.05 & er512-4 & 512 & 16 & 1 & 1.06 \\
    er256-5 & 256 & 16 & 2 & 1.05 & er512-5 & 512 & 16 & 2 & 1.06 \\
    er256-6 & 256 & 16 & 3 & 1.06 & er512-6 & 512 & 16 & 3 & 1.06 \\
    \bottomrule
  \end{tabular}
\end{table}

\begin{table}[tpb]
  \centering
  \caption{%
    Parameters and degree heterogeneity $\kappa$ of the random regular graphs (RRGs) used in our numerical analysis.
  }
  \label{tab.deghet:rrg.param}
  \begin{tabular}{lrrrr@{\hspace{12mm}}lrrrr}
    \toprule
    id & $N$ & $k$ & seed & $\kappa$ & id & $N$ & $k$ & seed & $\kappa$  \\
    \midrule
    rrg256-1 & 256 &  4 & 1 & 1.0 & rrg512-1 & 512 &  4 & 1 & 1.0 \\
    rrg256-2 & 256 &  4 & 2 & 1.0 & rrg512-2 & 512 &  4 & 2 & 1.0 \\
    rrg256-3 & 256 &  4 & 3 & 1.0 & rrg512-3 & 512 &  4 & 3 & 1.0 \\
    rrg256-4 & 256 & 16 & 1 & 1.0 & rrg512-4 & 512 & 16 & 1 & 1.0 \\
    rrg256-5 & 256 & 16 & 2 & 1.0 & rrg512-5 & 512 & 16 & 2 & 1.0 \\
    rrg256-6 & 256 & 16 & 3 & 1.0 & rrg512-6 & 512 & 16 & 3 & 1.0 \\
    \bottomrule
  \end{tabular}
\end{table}

\section{Annealed network approximation}
\label{app:ana}
The annealed network approximation replaces an entry of the adjacency matrix $A_{ij}$ with the expected number of edges from node $j$ to node $i$, i.e., $\aab{A_{ij}}$.
Here, we show that the common choice in Eq.~\eqref{eq:ana} preserves node degree while discarding degree correlations.

First, suppose that each node is characterized only by its degrees, $(\din{i}, \dout{i})$.
This is a simplification, because a node also has other characteristics, such as the identities of its neighbors.
Such local features are neglected in the annealed network approximation.
We write $\rmsub{P}{dest}(\din{}, \dout{} \mid \rmsub{\dout{}}{org})$ for the probability that an origin node whose outdegree is $\rmsub{\dout{}}{org}$ has an outgoing edge to one of the destination nodes with $(\din{}, \dout{})$.
Let $\hat{N}(\din{}, \dout{})$ denote the number of nodes with $(\din{}, \dout{})$.
Now consider the probability that an edge from node $j$, whose outdegree is $\dout{j}$, lands on node $i$, whose degrees are $(\din{i}, \dout{i})$.
It is given by
\begin{gather}
  \frac{\rmsub{P}{dest}\pab{\din{i}, \dout{i} \mid \dout{j}}}{\hat{N}\pab{\din{i}, \dout{i}}}.
\end{gather}
Using the joint degree distribution $P(\din{}, \dout{})$ and the number of nodes $N$, one expects, for large $N$,
\begin{gather}
  \hat{N}(\din{}, \dout{}) = N P(\din{}, \dout{}).
\end{gather}
Because node $j$ has $\dout{j}$ outgoing edges, we arrive at
\begin{gather} \label{eq:ana.corr}
  \aab{A_{ij}} = \dout{j} \frac{\rmsub{P}{dest}\pab{\din{i}, \dout{i} \mid \dout{j}}}{N P\pab{\din{i}, \dout{i}}}.
\end{gather}
The probability on the right-hand side is conditional on the outdegree of the origin node $j$, indicating degree correlations between nodes $i$ and $j$.

Next, we impose the no-correlation assumption.
Mathematically, it is expressed as
\begin{subequations} \label{eq:uncorr.assumpt}
\begin{align}
  \rmsub{P}{dest}\pab{\din{i}, \dout{i} \mid \dout{j}}
    =& \frac{\din{i}}{\sum_{l=1}^{N} \din{l}} \hat{N}(\din{i}, \dout{i})  \\
    =& \frac{\din{i}}{\aab{\din{}}} P(\din{i}, \dout{i}).
\end{align} \end{subequations}
The factor $\din{i} / \sum_l \din{l}$ in the first line denotes the probability that a randomly chosen edge of the network ends at node $i$.
The second equality is obtained by canceling $N$ in the numerator and the denominator.
We emphasize that the probability no longer depends on the outdegree of the origin, $\rmsup{d}{out}_j$, indicating the absence of degree correlations.
Substituting Eq.~\eqref{eq:uncorr.assumpt} into Eq.~\eqref{eq:ana.corr} yields
\begin{gather}
  \aab{A_{ij}} = \dout{j} \frac{\din{i}}{N \aab{\din{}}},
\end{gather}
which is equivalent to Eq.~\eqref{eq:ana}.
This choice preserves the degree of each node:
\begin{gather}
  \sum_{i=1}^N \aab{A_{ij}} = \dout{j},  \qquad
  \sum_{j=1}^N \aab{A_{ij}} = \din{i}.
\end{gather}

\section{Stochastic mean-field dynamics}
\label{app:smfd}
Here, we derive the networked extension of SMFD [Eq.~\eqref{eq:smfd}].
To simplify the notation, we define
\begin{gather}
  a(t) \coloneq K - f'(\Theta(t)).
\end{gather}
The evolution of $Z$ is, by definition,
\begin{subequations} \begin{align}
  \dot{Z} =& \sum_{j=1}^N 2 \rho_j y_j \dot{y}_j  \\
    =& \sum_{j=1}^N 2 \rho_j y_j \bab{-a y_j + \sqrt{2D} \hat{\eta}_j}  \\
    =& -2 a Z + 2 \sqrt{2D} \sum_{j=1}^N \rho_j y_j \hat{\eta}_j.
\end{align} \end{subequations}
Therefore, the dynamics of $Z$ are governed by the following SDE:
\begin{gather}
  \frac{\dot{Z}}{2} = -a(t) Z + \sqrt{2D} \hat{\eta}_Z,
\end{gather}
where we define the driving noise as
\begin{gather}
  \hat{\eta}_Z \coloneq \sum_{j=1}^N \rho_j y_j \hat{\eta}_j.
\end{gather}
At this point, one needs to keep track of all $y_j$ and $\hat{\eta}_j$ to compute the evolution of $Z$.
To obtain reduced dynamics, we approximate $\hat{\eta}_Z$ by another noise source that has statistical properties similar to those of $\hat{\eta}_Z$.

As preparation, we obtain the formal solution for $y_i$.
One can formally solve the OU process [Eq.~\eqref{eq:yi.dyn}] through a change of variables.
By setting the initial condition at $t = -\infty$, the influence of the initial condition is negligible, and we obtain
\begin{gather} \label{eq:yi.sol}
  y_i(t) = \frac{\sqrt{2D}}{\phi(t)} \int_{-\infty}^t \phi(t') \hat{\eta}_i(t') \d t',
\end{gather}
where we defined
\begin{gather}
  \phi(t) \coloneq \exp\pab{\int_0^t a(s) \d s}.
\end{gather}

We start by calculating the mean of $\hat{\eta}_Z$,
\begin{gather}
  \aab{\hat{\eta}_Z} = \sum_{j=1}^N \rho_j \aab{y_j \hat{\eta}_j}.
\end{gather}
Substituting the formal solution [Eq.~\eqref{eq:yi.sol}], one can proceed as follows:
\begin{subequations}
\begin{align}
  \aab{y_j \hat{\eta}_j} 
  =& \sqrt{2D} \aab{\frac{1}{\phi(t)} \int_{-\infty}^t \phi(s) \hat{\eta}_j(s) \hat{\eta}_j(t) \d s}  \\
  \approx& \frac{\sqrt{2D}}{\phi(t)} \int_{-\infty}^t \phi(s) \aab{\hat{\eta}_j(s) \hat{\eta}_j(t)} \d s  \\
  =& \frac{\sqrt{2D}}{\phi(t)} \pab{1 - 2 \rho_j + \frac{\kappa}{N}} \int_{-\infty}^t \phi(s) \delta(t - s) \d s  \\
  =& \frac{\sqrt{2D}}{2} \pab{1 - 2 \rho_j + \frac{\kappa}{N}}.
\end{align}
\end{subequations}
The approximation from the first to the second lines concerns the randomness of $\phi(t)$.
The correlation function of $\hat{\eta}_i$ [Eq.~\eqref{eq:corr.etahati}] was used to obtain the third line.
Technically, $\phi(t)$ is also a random variable affected by the fluctuation in $\Theta(t)$.
However, we assumed that its fluctuation was so small that one could consider $\phi(t)$ to be an external function of time that was independent of the focal randomness.
Using the result above, one finds that
\begin{gather}
  \aab{\hat{\eta}_Z} = \sqrt{\frac{D}{2}} \sum_{j=1}^N \rho_j \pab{1 - 2 \rho_j + \frac{\kappa}{N}}
    = \sqrt{\frac{D}{2}} \pab{1 - \frac{\kappa}{N}}.
\end{gather}

Next, we calculate the autocorrelation function,
\begin{gather}\label{eq:etaz.autocorr.orig}
\begin{split}
  & \aab[bigg]{\bab{\hat{\eta}_Z(s) - \aab{\hat{\eta}_Z(s)}} \bab{\hat{\eta}_Z(t) - \aab{\hat{\eta}_Z(t)}}} \\
  =& \aab{\hat{\eta}_Z(s) \hat{\eta}_Z(t)} - \aab{\hat{\eta}_Z(s)} \aab{\hat{\eta}_Z(t)}.
\end{split}
\end{gather}
One needs to calculate the first term:
\begin{subequations}
\begin{align}
  \begin{split}
  & \aab{\hat{\eta}_Z(s) \hat{\eta}_Z(t)}  \\
  =& \aab{\pab{\sum_{j=1}^N \rho_j y_j(s) \hat{\eta}_j(s)} \pab{\sum_{k=1}^N \rho_k y_k(t) \hat{\eta}_k(t)}}
  \end{split}  \\
  \label{eq:etaz.2ndmoment}
  =& \sum_{j=1}^N \sum_{k=1}^N \rho_j \rho_k \aab{y_j(s) \hat{\eta}_j(s) y_k(t) \hat{\eta}_k(t)}.
\end{align}
\end{subequations}
Substituting the formal solution [Eq.~\eqref{eq:yi.sol}], it follows that:
\begin{subequations}
\begin{align}
  \begin{split}
  & \aab{y_j(s) \hat{\eta}_j(s) y_k(t) \hat{\eta}_k(t)}  \\
  =& \biggg\langle
    \hat{\eta}_j(s) \hat{\eta}_k(t) \frac{2D}{\phi(s) \phi(t)}  \\
    & \qquad \pab{\int_{-\infty}^s \phi(s') \hat{\eta}_j(s') \d s'} \pab{\int_{-\infty}^t \phi(t') \hat{\eta}_k(t') \d t'}
  \biggg\rangle  
  \end{split}  \\
  \label{eq:smfd.deriv.int.corr}
  \begin{split}
  \approx& \frac{2D}{\phi(s) \phi(t)} \int_{-\infty}^s \d s' \bigg[  \\
  & \qquad \phi(s') \int_{-\infty}^t \d t' \phi(t') \aab{\hat{\eta}_j(s') \hat{\eta}_j(s) \hat{\eta}_k(t') \hat{\eta}_k(t)} \bigg].
  \end{split}
\end{align}
\end{subequations}
Again, we treated $\phi(t)$ as if it were an external function of time in the last line.
Since $\hat{\eta}_j$ are Gaussian random variables, their fourth order moment can be decomposed into the sum of products of pairwise correlation functions~\cite[Eq.~(2.8.6)]{gardiner2009Stochastic}:
\begin{subequations} \begin{align} \begin{split}
  & \aab{\hat{\eta}_j(s') \hat{\eta}_j(s) \hat{\eta}_k(t') \hat{\eta}_k(t)}  \\
  =& \aab{\hat{\eta}_j(s') \hat{\eta}_j(s)} \aab{\hat{\eta}_k(t') \hat{\eta}_k(t)}  \\
    & + \aab{\hat{\eta}_j(s') \hat{\eta}_k(t')} \aab{\hat{\eta}_j(s) \hat{\eta}_k(t)}  \\
    & + \aab{\hat{\eta}_j(s') \hat{\eta}_k(t)} \aab{\hat{\eta}_j(s) \hat{\eta}_k(t')}
  \end{split}  \\
  \begin{split}
  =& \pab{1 - 2 \rho_j + \frac{\kappa}{N}} \pab{1 - 2 \rho_k + \frac{\kappa}{N}} \delta(s - s') \delta(t - t')  \\
  & + \pab{\delta_{jk} - \rho_j - \rho_k + \frac{\kappa}{N}}^2  \\
  & \qquad \bab[bigg]{\delta(t' - s') \delta(t - s) + \delta(t - s') \delta(t' - s)}.
  \end{split}
\end{align} \end{subequations}
Therefore, assuming $s \leq t$, the integral in Eq.~\eqref{eq:smfd.deriv.int.corr} becomes
\begin{gather}
\begin{split}
  & \int_{-\infty}^s \d s' \phi(s') \int_{-\infty}^t \d t' \phi(t') \aab{\hat{\eta}_j(s') \hat{\eta}_j(s) \hat{\eta}_k(t') \hat{\eta}_k(t)}  \\
  =& \frac{\phi(s) \phi(t)}{4} \pab{1 - 2 \rho_j + \frac{\kappa}{N}} \pab{1 - 2 \rho_k + \frac{\kappa}{N}}  \\
  & \quad + \delta(t - s) \pab{\delta_{jk} - \rho_j - \rho_k + \frac{\kappa}{N}}^2 \int_{-\infty}^s \phi(s')^2 \d s'.
\end{split}
\end{gather}
Thus, Eq.~\eqref{eq:etaz.2ndmoment} is approximately
\begin{gather} \label{eq:etaz.2ndmoment.deriv1}
  \begin{split}
  & \aab{\hat{\eta}_Z(s) \hat{\eta}_Z(t)}  \\
  \approx& \frac{D}{2} \sum_{j=1}^N \sum_{k=1}^N \rho_j \rho_k \pab{1 - 2 \rho_j + \frac{\kappa}{N}} \pab{1 - 2 \rho_k + \frac{\kappa}{N}}  \\
  & + \frac{2D}{\phi(s) \phi(t)} \delta(t - s)  \\
  & \qquad \sum_{j=1}^N \sum_{k=1}^N
    \rho_j \rho_k \pab{\delta_{jk} - \rho_j - \rho_k + \frac{\kappa}{N}}^2 \int_{-\infty}^s \phi(s')^2 \d s'.
  \end{split}
\end{gather}
The first term simplifies as follows:
\begin{subequations} \begin{align}
\begin{split} 
  & \frac{D}{2} \sum_{j=1}^N \sum_{k=1}^N \rho_j \rho_k \pab{1 - 2 \rho_j + \frac{\kappa}{N}} \pab{1 - 2 \rho_k + \frac{\kappa}{N}}  \\
  =& \frac{D}{2} \pab{1 - \frac{\kappa}{N}}^2
  \end{split}  \\
  =& \aab{\hat{\eta}_Z}^2.
\end{align} \end{subequations}
In the second term, we calculate the part involving summations:
\begin{subequations} \begin{align}
  \begin{split}
  & \sum_{j,k} \rho_j \rho_k \pab{\delta_{jk} - \rho_j - \rho_k + \frac{\kappa}{N}}^2  \\
  =& \sum_{j,k} \rho_j \rho_k \bigg[
    \delta_{jk} \pab{1 + \frac{2 \kappa}{N}} + \pab{\frac{\kappa}{N}}^2 \\
    & \quad + \rho_j^2 + 2 \rho_j \rho_k + \rho_k^2 - 2 \delta_{jk} \pab{\rho_j + \rho_k} - \frac{2\kappa}{N} \pab{\rho_j + \rho_k}
  \bigg]
  \end{split}  \\
  \begin{split}
  =& \frac{\kappa}{N} \pab{1 + \frac{2\kappa}{N}} + \pab{\frac{\kappa}{N}}^2 + 2 \sum_{j=1}^N \rho_j^3  \\
  & + 2 \pab{\frac{\kappa}{N}}^2 - 4 \sum_{j=1}^N \rho_j^3 - \pab{\frac{2\kappa}{N}}^2  
  \end{split}  \\
  =& \frac{\kappa}{N} - \frac{2 \kappa_3}{N^2} + \pab{\frac{\kappa}{N}}^2,
\end{align} \end{subequations}
where we used $\kappa_3$ defined by
\begin{gather} \label{eq:kappa3}
  \kappa_3 \coloneq \frac{\aab{\pab{\dout{}}^3}}{\aab{\pab{\dout{}}}^3}.
\end{gather}
At the second equality, we used the following relations:
\begin{subequations} \begin{align}
  &\sum_{j,k} \rho_j \rho_k = \pab{\sum_j \rho_j} \pab{\sum_k \rho_k} = 1,  \\
  &\sum_{j,k} \delta_{jk} \rho_j \rho_k = \sum_j \rho_j^2 = \frac{\kappa}{N},  \\
  &\sum_{j,k} \rho_j^2 \rho_k = \sum_{j,k} \rho_j \rho_k^2 = \pab{\sum_j \rho_j^2} \pab{\sum_k \rho_k} = \frac{\kappa}{N},  \\
  &\sum_j \rho_j^3 = \sum_j \pab{\frac{\rmsup{d}{out}_j}{N \aab{\rmsup{d}{out}}}}^3 = \frac{1}{N^2} \frac{\aab[big]{\pab[big]{\rmsup{d}{out}}^3}}{\aab{\rmsup{d}{out}}^3} = \frac{2 \kappa_3}{N^2},  \\
  &\sum_{j,k} \rho_j^2 \rho_k^2 = \pab{\sum_j \rho_j^2} \pab{\sum_k \rho_k^2} = \pab{\frac{\kappa}{N}}^2.
\end{align} \end{subequations}
Consequently, the second term of Eq.~\eqref{eq:etaz.2ndmoment.deriv1} simplifies to
\begin{gather}
  2D \delta(t - s) \bab{\frac{\kappa}{N} - \frac{2 \kappa_3}{N^2} + \pab{\frac{\kappa}{N}}^2} \int_{-\infty}^s \bab{\frac{\phi(s')}{\phi(s)}}^2 \d s'.
\end{gather}
Let us define
\begin{gather}
  D_Z(t) \coloneq \int_{-\infty}^t \bab{\frac{\phi(s)}{\phi(t)}}^2 \d s,
\end{gather}
which is a solution of the following differential equation:
\begin{gather}
  \dot{D}_Z(t) = 1 - 2 a(t) D_Z(t).
\end{gather}
Put together, Eq.~\eqref{eq:etaz.2ndmoment} becomes
\begin{gather} \begin{split}
  & \aab{\hat{\eta}_Z(t) \hat{\eta}_Z(t+\tau)}  \\
  =& \aab{\hat{\eta}_Z}^2 + \delta(\tau) 2D \bab{\frac{\kappa}{N} - \frac{2 \kappa_3}{N^2} + \pab{\frac{\kappa}{N}}^2} D_Z(t).
\end{split} \end{gather}
Coming back to Eq.~\eqref{eq:etaz.autocorr.orig}, we finally obtain the autocorrelation function:
\begin{gather}
\begin{split}
  & \aab{\bab[bigg]{\hat{\eta}_Z(t) - \aab{\hat{\eta}_Z(t)}} \bab[bigg]{\hat{\eta}_Z(t+\tau) - \aab{\hat{\eta}_Z(t+\tau)}}} \\
  =& \delta(\tau) 2D \bab{\frac{\kappa}{N} - \frac{2 \kappa_3}{N^2} + \pab{\frac{\kappa}{N}}^2} D_Z(t).
\end{split}
\end{gather}

Another statistical property of interest is the correlation between $\hat{\eta}_\Theta$ and $\hat{\eta}_Z$.
Their cross-correlation is:
\begin{subequations}
\begin{align}
  \begin{split}
  & \aab{\hat{\eta}_\Theta(t) \hat{\eta}_Z(t+\tau)}  \\
  =& \aab{\hat{\eta}_\Theta(t) \bab{\sum_{j=1}^N \rho_j y_j(t + \tau) \hat{\eta}_j(t+\tau)}}  
  \end{split}  \\
  \begin{split}
  =& \sqrt{2D} \sum_{j=1}^N \rho_j  \\
  & \quad \aab{\hat{\eta}_\Theta(t) \bab{\frac{1}{\phi(t+\tau)} \int_{-\infty}^{t+\tau} \phi(s) \hat{\eta}_j(s) \d s} \hat{\eta}_j(t+\tau)}
  \end{split}  \\
  \approx& \frac{\sqrt{2D}}{\phi(t+\tau)} \sum_{j=1}^N \rho_j \int_{-\infty}^{t+\tau} \phi(s) \aab{\hat{\eta}_\Theta(t) \hat{\eta}_j(s) \hat{\eta}_j(t+\tau)} \d s.
\end{align}
\end{subequations}
We used the same assumption as above in the approximation of the third line.
As $\hat{\eta}_\Theta$ and $\hat{\eta}_j$ are Gaussian random variables whose means are zero, odd moments among them vanish~\cite[Sect.~2.8.1]{gardiner2009Stochastic}.
It follows that
\begin{gather}
  \aab{\hat{\eta}_\Theta(t) \hat{\eta}_j(s) \hat{\eta}_j(t+\tau)} \equiv 0,
\end{gather}
and hence there is no correlation between $\hat{\eta}_\Theta$ and $\hat{\eta}_Z$:
\begin{gather}
  \aab{\hat{\eta}_\Theta(t) \hat{\eta}_Z(t+\tau)} \equiv 0.
\end{gather}

Finally, in light of the results above, we approximate $\hat{\eta}_Z$ by a white Gaussian noise.
With a white Gaussian noise $\eta_Z(t)$ satisfying
\begin{subequations} \begin{align}
  & \aab{\eta_Z(t)} \equiv 0, \\
  & \aab{\eta_Z(t) \eta_Z(t + \tau)} = \delta(\tau),  \\
  & \aab{\eta_Z(t) \eta_\Theta(t+\tau)} \equiv 0,
\end{align} \end{subequations}
we approximate $\hat{\eta}_Z(t)$ as
\begin{gather}
\begin{split}
  \hat{\eta}_Z(t) \approx& \sqrt{\frac{D}{2}} \pab{1 - \frac{\kappa}{N}} \\
  & + \sqrt{2D} \sqrt{\frac{\kappa}{N} - \frac{2 \kappa_3}{N^2} + \pab{\frac{\kappa}{N}}^2} \sqrt{D_Z(t)} \, \eta_Z(t).
\end{split}
\end{gather}
Consequently, one obtains the following system of SDEs, which approximates the evolution of $\Theta$:
\begin{subequations} \label{eq:smfd.system}
\begin{align}
  \dot{\Theta} =& f(\Theta) + \frac{f''(\Theta)}{2} Z + \sqrt{2D} \sqrt{\frac{\kappa}{N}} \eta_\Theta,  \\
  \begin{split}
    \frac{\dot{Z}}{2} =& -\bab{K - f'(\Theta)} Z + D \pab{1 - \frac{\kappa}{N}}  \\
      & + 2D \sqrt{\frac{\kappa}{N} - \frac{2 \kappa_3}{N^2} + \pab{\frac{\kappa}{N}}^2} \sqrt{D_Z} \eta_Z,
  \end{split}  \\
  \dot{D}_Z =& 1 - 2 \bab{K - f'(\Theta)} D_Z,
\end{align}
\end{subequations}
where $\eta_\Theta$ and $\eta_Z$ satisfy
\begin{subequations} \begin{align}
  & \aab{\eta_\Theta(t)} = 0,  \quad
  \aab{\eta_\Theta(t) \eta_\Theta(t+\tau)} = \delta(\tau),  \\
  & \aab{\eta_Z(t)} = 0,  \quad
  \aab{\eta_Z(t) \eta_Z(t+\tau)} = \delta(\tau),  \\
  & \aab{\eta_Z(t) \eta_\Theta(t+\tau)} = 0.
\end{align} \end{subequations}
This approximation preserves the mean, autocorrelation, and cross-correlation with $\hat{\eta}_\Theta$ of the original $\hat{\eta}_Z$.
Neglected are higher-order moments of $\hat{\eta}_Z$.
Indeed, it is not obvious whether $\hat{\eta}_Z(t)$ actually is Gaussian.
In addition to $\hat{\eta}_j$, $y_j$ is also Gaussian after the relaxation to the steady state.
However, since $\hat{\eta}_Z$ is defined as the sum of products of two Gaussian random variables $\hat{\eta}_j$ and $y_j$, identifying its distribution is not a trivial task in general.

The derived system [Eq.~\eqref{eq:smfd.system}] indicates that, for sufficiently large $K$ satisfying $K \gg f(\Theta)$ and $K \gg f'(\Theta)$, $Z$ and $D_Z$ evolve much faster than $\Theta$.
In other words, $Z$ and $D_Z$ evolve on much faster time scales than that of $\Theta$.
Accordingly, we assume that $Z$ and $D_Z$ always assume their stationary values so long as the $\Theta$ dynamics are concerned, which is the technique known as ``adiabatic elimination''.
Solving $\dot{D}_Z = 0$ for $D_Z$, one finds its stable fixed point, which is
\begin{gather}
  D_Z^* = \frac{1}{2\bab{K - f'(\Theta)}}.
\end{gather}
Substitution of $D_Z^*$ into the $Z$ dynamics yields
\begin{gather}
\begin{split}
  \frac{\dot{Z}}{2} =& -\bab{K - f'(\Theta)} Z + D \pab{1 - \frac{\kappa}{N}}  \\
    & + \frac{\sqrt{2} D}{\sqrt{K - f'(\Theta)}} \sqrt{\frac{\kappa}{N} - \frac{2 \kappa_3}{N^2} + \pab{\frac{\kappa}{N}}^2} \eta_Z,
\end{split}
\end{gather}
The noise intensity is quantified by the diffusion coefficient of the corresponding FPE, which is the squared prefactor of a white Gaussian noise.
The diffusion coefficient of the $\Theta$ dynamics is
\begin{gather}
  \frac{2D \kappa}{N} = \order{\frac{D \kappa}{N}},
\end{gather}
whereas that of the $Z$ dynamics is
\begin{gather}
  \frac{2D^2}{K - f'(\Theta)} \bab{\frac{\kappa}{N} - \frac{2\kappa_3}{N^2} + \pab{\frac{\kappa}{N}}^2} = \order{\frac{D^2 \kappa}{NK}}.
\end{gather}
Therefore, when $D \ll K$, the fluctuation in $Z$ is negligible compared to that of $\Theta$.
The noise-free dynamics of $Z$ are governed by
\begin{gather}
  \frac{\dot{Z}}{2} = -\bab{K - f'(\Theta)} Z + D \pab{1 - \frac{\kappa}{N}},
\end{gather}
whose stable fixed point is $Z^*(\Theta)$ defined in Eq.~\eqref{eq:smfd.zstar} of the main text.
Substituting $Z^*$ into the $\Theta$ dynamics of the SDE-system [Eq.~\eqref{eq:smfd.system}] yields the one-dimensional reduced dynamics of $\Theta$ shown in the main text [Eq.~\eqref{eq:smfd}].

The associated potential of the reduced $\Theta$ dynamics [Eq.~\eqref{eq:smfd.theta}] is
\begin{gather} \begin{split}
  V_\Theta(\Theta)
  =& \frac14 \Theta^4 - \frac{1 + r}{3} \Theta^3  \\
  & + \frac{r + 3Z^*(\Theta)}{2} \Theta^2 - \pab{1 + r} Z^*(\Theta) \Theta.
\end{split} \end{gather}
Substituting $V_\Theta$ and adjusting the diffusion coefficient appropriately, the mean first passage time formula [Eq.~\eqref{eq:mfpt}] becomes
\begin{gather} \label{eq:smfd.met} \begin{split}
  & \rmsub{T}{SMFD}\pab{K, \frac{\kappa}{N}}  \\
  =& \frac{N}{\kappa D} \int_0^\xi \d y \int_{-\infty}^y \d z \exp\pab{-\frac{N}{\kappa D} \bab{V_\Theta(y) - V_\Theta(z)}}.
\end{split} \end{gather}

\section{Numerical methods}
\label{app:num.methods}

All numerical analysis was performed in Julia using Docker containers, as well as Apptainer containers on a high-performance computer.
Figures were generated using \texttt{CairoMakie.jl}, a backend of \texttt{Makie.jl}~\cite{danisch2021Makiejl}.

Network instances on which the original networked model [Eq.~\eqref{eq:original.model}] was simulated were generated using \texttt{Graphs.jl}~\cite{fairbanks2021JuliaGraphs}.
In particular, we generated networks using the following function calls:
\begin{itemize}
  \item \texttt{complete\_bipartite\_graph($N-n_2$, $n_2$)} with parameter values listed in Tab.~\ref{tab.deghet:cbg.param}.
  \item \texttt{static\_scale\_free($N$, $m N$, $\gamma$; seed=seed)} with parameters in Tab.~\ref{tab.deghet:ssf.param}. The parameter $m$ roughly corresponds to half the mean degree, and the exponent of the power law is $\gamma$.
  \item \texttt{erdos\_renyi($N$, $m N / 2$; seed=seed)} with parameters in Tab.~\ref{tab.deghet:er.param}. The parameter $m$ roughly corresponds to the mean degree.
  \item \texttt{random\_regular\_graph($N$, $k$; seed=seed)} with parameters in Tab.~\ref{tab.deghet:rrg.param}. The parameter $k$ is the degree.
\end{itemize}
Upon network generation, we ensured that each network was connected by extracting the largest connected component, if necessary.

In a standard approach to numerical integration of dynamical systems on a network, one would first generate a network instance, obtain its adjacency or Laplacian matrix, implement the focal system using the matrix, and then solve an initial value problem using a numerical solver.
Sparse matrix implementations are often helpful, since the computation involves an $N \times N$ matrix that is sparse in many cases.
In this work, instead of constructing a Laplacian matrix ourselves, we used a Julia package, \texttt{NetworkDynamics.jl}~\cite{lindner2021NetworkDynamicsjlComposing}, to simulate the networked SDEs.
An escape of each element, defined as an upward zero-crossing event of $x_i - \xi$, was detected using callback functionality, specifically \texttt{SciMLBase.ContinuousCallback}.

To predict mean escape times based on NlinMFFPE, we adopted the same method as in the companion Letter~\jointarticle{}.
That is, we first solved a discretized version of NlinMFFPE on the interval $x \in (a, b)$.
Spatial derivatives were discretized with the step size $\Delta x$ using a central difference scheme.
The resulting ODE system was solved using a variant of the Runge--Kutta method, \texttt{Tsit5()}~\cite{tsitouras2011Runge}, in \texttt{DifferentialEquations.jl}~\cite{rackauckas2017Differentialequationsjla}.
The numerical integration was terminated when the mean field $X$ approached the active state (specifically, $X > 0.9$) and the probability current at the threshold, $J(\xi, t)$, became sufficiently small to satisfy $J(\xi, t) \leq J_{\xi,\mathrm{min}}$.
The termination was implemented using callback functionality.
Results of numerical integration were recorded at time intervals of $\Delta t$.
Specific values of numerical parameters are reported in the relevant figures to eliminate room for human error in transcription.
The integral was approximated by a summation computed from the resulting time series of $J(\xi, t)$.
To evaluate the mean first passage time formula [Eq.~\eqref{eq:mfpt}], we used \texttt{Integrals.jl}.
The bifurcation point of DMFD, $K_2$, was calculated using \texttt{find\_zero()} in \texttt{Roots.jl}.

\bibliography{25-phd-thesis_deghet}

\end{document}